\documentclass{appolb}
\usepackage{epsfig}

  \newcommand{\kpen}{\ensuremath{K_{e3}}}
\newcommand{\Kpeng}
  {\ensuremath{K_{L} \rightarrow \pi^{\pm} e^{\mp} \nu_{e} \gamma}}
  \newcommand{\kpeng}{\ensuremath{K_{e3\gamma}}}

  \newcommand{\kpmz}{\ensuremath{K_{\pi3}}}

\newcommand{\ChPT} {\ensuremath{\rm\chi PT}}

\newcommand{\Acc}{\ensuremath{\rm Acc}}
\newcommand{\BR}{\ensuremath{\rm BR}}
\newcommand{\Num}{\ensuremath{\rm N}}

\newcommand{\bcm}{\left(\begin{array}{c}}
\newcommand{\ecm}{\end{array}\right)}
\newcommand{\bccm}{\left(\begin{array}{cc}}
\newcommand{\eccm}{\end{array}\right)}
\begin{document}
%\date{\today}
\pagestyle{plain}
\newcount\eLiNe\eLiNe=\inputlineno\advance\eLiNe by -1
\title{A New Measurement of the Radiative \kpen\ Branching Ratio and
  Photon Spectrum}
\author{Douglas R Bergman\\for the KTeV Collaboration
\address{Department of Physics and Astronomy\\
  Rutgers University\\
  136 Frelinghuysen Rd\\
  Piscataway NJ 08854, USA}}
\maketitle

\begin{abstract}
  We present a preliminary report on a new measurement of the
  radiative \kpen branching ratio and the first study of the photon
  spectrum in this decay.  We find $\BR(\kpeng, E_\gamma^*>30{\rm\
  GeV}, \theta_{e\gamma}^*>20^\circ)/\BR(\kpen) = 0.911
  \pm0.009\textrm{(stat.)}^{+0.021}_{-0.010}\textrm{(syst.)}\%$.  Our
  measurement of the spectrum is consistent with inner bremsstrahlung
  only as the source of photons at the $2\sigma$ level.
\end{abstract}

A good understanding of radiative \kpen\ (\Kpeng) decays is important
for many analyses in precision neutral kaon experiments.  The decay is
interesting in its own right as a test of Current Algebra (CA) and
Chiral Perturbation Theory (\ChPT) models of kaon structure.  These
models predict the size of the direct emission (DE) component, where
the photon comes directly from the decay vertex, with respect to the
dominant internal bremsstrahlung (IB) component, where the photon
comes from one of the external particles.  CA predicts a DE component
of about 1\%\cite{FFS:1970}, whereas \ChPT\ predicts a smaller DE component
at the tenth of a percent level\cite{Bijnens:1992}.

\begin{figure} 
\begin{center}
\epsfxsize=3in
\epsfbox{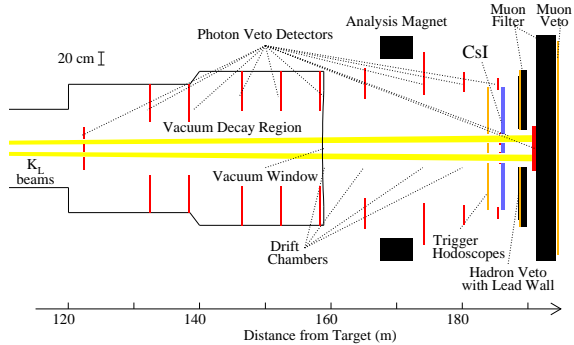}
\caption{The KTeV Detector}
\label{fig:ktev} 
\end{center}
\end{figure}

The measurement described here was preformed using the KTeV detector
at Fermilab, shown in Figure~\ref{fig:ktev}, which was designed to
measure $\epsilon'/\epsilon$.  Two nearly parallel beams of $K^0$ are
produced by the 800 GeV Tevatron proton beam on a Beryllium target,
with collimators well upstream of the decay region allowing the $K_S$
component to die away.  (The $K_S$ regenerator used in
$\epsilon'/\epsilon$ was removed for this data.)  Decay products are
analyzed with a spectrometer consisting of four drift chambers and an
analysis magnet.  Following the spectrometer is a pure CsI
electromagnetic calorimeter and a muon filter.

The strategy used in the analysis is to identify an inclusive \kpen\ 
sample, then identify a subsample containing one photon.  The \kpen\ 
sample is defined by requiring exactly two tracks in the spectrometer.
The X and Y projections of the tracks are matched using clusters in
the calorimeter, implying one cluster per track.  Particle
identification is performed by comparing the momentum in the
spectrometer with the energy in the calorimeter.  The electron must
have $0.95<E/P<1.1$, the pion, $E/P<0.7$.  The momentum of each track
must be above 7 GeV to make the muon filter efficient, and hits in the
muon filter, or any other veto detector, kill the event.
Kinematically disallowed events are removed as well.  Fiducial cuts
are placed on all tracks and clusters to restrict events to well
understood parts of the detector.  \kpmz\ background events are
eliminated by removing kinematically allowed event under the \kpmz\ 
ansatz.  The two-fold ambiguity in the reconstructed kaon momentum is
resolved by taking the more likely solution for a given pair using MC
generated distributions for allowed solution pairs.

With the inclusive \kpen\ sample in hand, a subsample of events with
one photon candidate is isolated.  A photon candidate appears as a
cluster in the calorimeter unconnected with a track.  The cluster must
be more that 8 cm away from the electron and 40 cm away from the pion,
to avoid fake extra clusters caused by shower fluctuations, which for
pions especially, are poorly understood.  The cluster must have an
energy of 3 GeV, and pass a shape $\chi^2$ cut for being a cluster
from a single, electromagnetic shower.  There must be only one cluster
satisfying a looser set of candidate cuts (30 cm from the pion and 1
GeV on the energy).  The difference in the definition of clusters for
the final photon sample and for vetoing an event leads to a
correction, as the MC veto rate is lower than observed in data.

The data for this sample was taken at a low rate reducing the problem
of accidental photons.  The effect is still important however, so an
accidental trigger was used to sample the state of the detector at
arbitrary times.  These events were overlayed on MC generated events
to accurately model the data.  In all, accidentals contribute about
4\% to the photonic subsample.

It has become customary to cut on the photon energy in the kaon center
of momentum frame (CM) and on the angle between the photon and the
electron in that frame.  The first cut is necessary since the
radiative branching ratio is IR divergent in the photon energy, yet
experiments are always limited by detection efficiency to some minimum
energy.  The second cut removes external bremsstrahlung radiation
which is very highly correlated with the electron direction.  Standard
cuts at 30 MeV and 20 degrees are used, though KTeV is sensitive to
significantly lower values in both cases.

The measured branching ratio is simply defined
\begin{eqnarray*}
        \frac{\Gamma(\kpeng,
                E_\gamma^*>30{\rm\ GeV},
                \theta_{e\gamma}^*>20^\circ)}
             {\Gamma(\kpen)} 
             &=& \frac{\Num(\kpeng)}{\Num(\kpen)}
                        \frac{\Acc(\kpen)}{\Acc(\kpeng)} \\ 
             &=&   0.911\pm0.009({\rm stat})^{+0.021}_{-0.011}({\rm syst})\%
\end{eqnarray*}
where the samples used in the calculation are as follows
\[
\begin{array}{|l|rrrl|}\hline
  & \textrm{Data}& \textrm{MC Gen}& \textrm{MC Anal}&\textrm{Accept}\\ \hline
\kpen \textrm{, no BG sub}        &5760880&187408000&19979000&0.106612(23)\\
\kpen \textrm{, BG sub}           &5760140&&&\\\hline
\kpeng \textrm{, (20,30), no BG sub}&15575&  1750190&   55262&0.03158(13)\\
\kpeng \textrm{, (20,30), veto corr}&     &         &   55848&0.03191(13)\\
\kpeng \textrm{, (20,30), BG sub}   &15379&&&\\\hline
\end{array}
\]
Systematic uncertainties come from disagreements various data and MC
distributions (0.004\%), variation of the answer with the value of the
$E_\gamma$ cut ($^{+0.019}_{-0.005}$\%), the uncertainty in the \kpen\ 
form factor (0.007\%) and the uncertainty on the veto correction
(0.005\%) for a total systematic uncertainty of
$^{+0.021}_{-0.010}$\%.  The veto correction is small, 1\%, but known
only to 20\% of itself. 

\begin{figure}
\begin{center}
\epsfysize=2in
\epsfbox{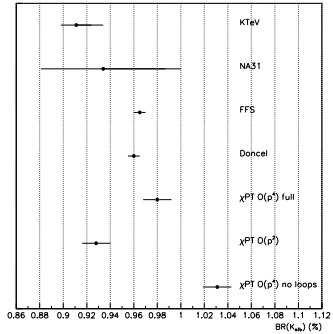}
\end{center}
\caption{A comparison of the BR result with other recent measurements and
predictions.}
\label{fig:brcomp}
\end{figure}

The observed branching ratio is consistent with the previous
measurement by NA31\cite{Leber:1996}, while considerably lower than both
the CA (FFS\cite{FFS:1970} and Doncel\cite{Doncel:1970})
and \ChPT\ predictions, as shown in Figure~\ref{fig:brcomp}.

Fearing \textit{et al.}\cite{FFS:1970} give a phenomenological model of
DE which was used to study the photon energy spectrum.  They give the
matrix element for \kpeng\ in the soft-kaon approximation as
\begin{eqnarray*}
T(\kpeng) &=& T_{\rm IB}\\
&&      +\frac{C}{M^2}(\epsilon\cdot l Q\cdot k - \epsilon\cdot Q l\cdot k) +
        \frac{D}{M^2}(\epsilon_{\mu\nu\alpha\beta}
                        \epsilon^\mu l^\nu Q^\alpha k^\beta)
\end{eqnarray*}
where $M$ is the kaon mass, $\epsilon$ is photon polarization, $l$ is
the electron-neutrino current vector, $k$ is the photon momentum and
$Q$ is the pion momentum.  Terms ($A$ and $B$) with the kaon momentum
replacing the pion momentum been left out in this approximation.

The photon spectrum was generated at a lattice of points in $CD$ space
and compared to the acceptance-corrected spectrum seen in the data.
The comparisons used photons with CM energy between 25 MeV and 200 MeV
and more than 5 degrees in the CM from the electron.  The lattice of
comparison points is not aligned with the $CD$ axes so the $\chi^2$
values of the comparisons were fit to forth degree polynomials along
rows, and the resulting fit parameters were themselves fit in the
orthogonal direction.  These fits serve as an interpolation between
the measured points and allow the generation of $\sigma$ contours in
the $CD$ plane.  

\begin{figure}
\begin{center}
\epsfysize=2in
\epsfbox{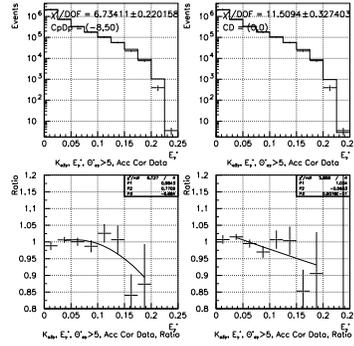}
\end{center}
\caption{Data/MC comparisons of the $E_\gamma^*$ spectrum at the best
$C'D'$ point and at the IB-only point.}
\label{fig:gcdppqtfch2}
\end{figure}

The photon spectrum comparisons at two lattice points, the best point
(left) and the IB-only point (right), are shown in
Figure~\ref{fig:gcdppqtfch2}.  The most significant difference between
the two is a slight hardening of the photon spectrum in the 50--100
MeV region.

\begin{figure}
\begin{center}
\epsfysize=2in
\epsfbox{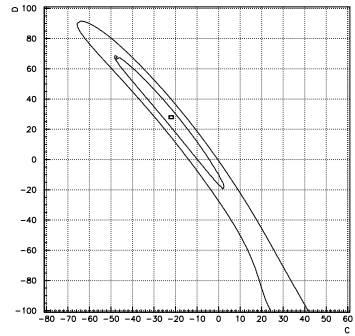}
\end{center}
\caption{The first two constant $\sigma$ contours of the
fit/interpolated $\chi^2$ surface as a function of $CD$.  The box
shows the bin the lowest value.}
\label{fig:gcdppqtfcd}
\end{figure}

The 1$\sigma$ and 2$\sigma$ contours generated from interpolation in
the $CD$ frame are shown in Figure~\ref{fig:gcdppqtfcd}.  The contours
are completely contained within the simulated domain.  The result is
different from the IB-only spectrum at the 2$\sigma$ level, which is
too small to allow one to claim to have seen DE in this mode, but is
enticingly similar to the difference between the branching ratio
measurement and theoretical predictions.

\begin{thebibliography}{10}

\bibitem{FFS:1970}
H.~Fearing et al.
\newblock Phys. Rev. D {bf 2} (1970) 542.

\bibitem{Doncel:1970}
M.~G.~Doncel
\newblock Phys. Lett. {\bf 32B} (1970) 623.

\bibitem{Bijnens:1992}
J.~Bijnens et al.
\newblock Semileptonic kaon decays in chiral perturbation theory.
\newblock hep-ph/9208204.

\bibitem{Leber:1996}
F.~Leber et al. 
\newblock Phys. Lett. B {\bf 369} (1996) 69.

\end{thebibliography}
\end{document}